\documentclass[10pt]{article}
\usepackage[utf8]{inputenc}
\usepackage{amsmath,amsfonts,amssymb,latexsym}
\usepackage[T1]{fontenc}
\newtheorem{satz}{Theorem}[section]

\newtheorem{assumption}[satz]{Assumption}
\newtheorem{defi}[satz]{Definition}

\newtheorem{bem}[satz]{Remark}
\newtheorem{lemma}[satz]{Lemma}
\newtheorem{koro}[satz]{Corollary}
\newtheorem{conclusion}[satz]{Conclusion}
\newtheorem{ob}[satz]{Observation}

\newtheorem{conjecture}[satz]{Conjecture}

\newcommand{\mcal}{\mathcal}
\newcommand{\mbf}{\mathbf}

\newcommand{\tit}{\textit}

\newcommand{\N}{\mathbb{N}}

\newcommand{\beq}{\begin{equation}}
\newcommand{\eeq}{\end{equation}}

\begin{document}
\thispagestyle{empty}
\begin{center}
\vspace*{1.0cm}
{\Large{\bf The Thermal Substructure of General Relativity}}
\vskip 1.5cm
 
{\large{\bf Manfred Requardt}}

\vskip 0.5cm

Institut fuer Theoretische Physik\\
Universitaet Goettingen\\
Friedrich-Hund-Platz 1\\
37077 Goettingen \quad Germany\\
(E-mail: requardt@theorie.physik.uni-goettingen.de or muw.requardt@googlemail.com) 

\end{center}
\begin{abstract}
In a first step we will provide arguments for the understanding of quantum space-time (QST), that means, the microscopic substructure which is assumed to underly ordinary smooth classical
space-time, as a thermal system at each (macroscopic) point $x$ of the classical space-time manifold (ST). In this context we exploit among other things some recent findings in the foundations of quantum statistical mechanics. 

We argue that the classical metric tensor field $g_{ij}(x)$ plays the role of an order parameter field, signalling the existence of a primordial phase transition in which both space-time and (quantum) matter emerged, accompanied by the spontaneous symmetry breaking (SSB) of diffeomorphism invariance which is induced by the existence of quantum excitations in QST. 

We then analyze the role the Einstein equations (EEQ) are playing in this context and relate them to the local thermodynamical heat influx into QST at point $x$, thus defining the Einstein tensor or Ricci tensor as local heat tensor. We also define local gravitational temperature, entropy and internal energy.

Finally, we give a thermodynamic interpretation of the conserved Noether current which derives from diffeomorphism invariance, separating it into heat contributions and work terms. In this connection we discuss the role and nature of tensor and non-tensor quantities and relate our investigation to the socalled energy vector of Hilbert`s classical analysis.

\end{abstract}
\newpage
\section{Introduction}
Without question one of the most unexpected developements in general relativity was the realisation that there exists a close relationship between certain laws of classical black hole (BH) physics and the laws of thermodynamics (\cite{Bard1}). This fundamental connection was first described by Bekenstein (\cite{Be1},\cite{Be2}) and a little bit later supported by the fundamental analysis of Hawking (\cite{Haw1}). A nice overview is given for example in \cite{Wald1}.

That there might perhaps exist a closer and even more fundamental connection between classical relativity and thermodynamics was, as far as we know, emphasized by Jacobson (\cite{Jacob1}) and further developed in e.g. \cite{Jacob2} and many papers of Padmanabhan (see, for example,\cite{Pad1},\cite{Pad2} or \cite{Pad3}). It was shown that one can infer from the physics near horizons, in particular the event horizon of a black hole, that the Einstein equation (EEQ) must hold. One important tool in this context was the use of approximate Rindler-Unruh horizons and the corresponding results of Fulling-Unruh (\cite{Full1},\cite{Un1}).

One should remark that, as far as we can see, all the many papers which do exist in this context emphasize the importance of horizons and the role which is played by the \tit{entanglement entropy} of systems behind these horizons. That is, from this view point, entropy and more generally thermodynamics appear to be essentially located near these horizons and not e.g. in the bulk. These properties seem to be suggested, for example, by the famous \tit{entropy area law} of black holes (BH).

Whereas horizons apparently seem to play such a fundamental role for the thermal behavior of gravitation, we want to go a step further and will argue in this paper that in our view the situation is a different one and that both the Einstein equation (EEQ) and gravitation in general are of a thermal nature being independent of the existence of horizons and which can be inferred by analyzing the ordinary bulk situation in general relativity (GR). The special behavior near horizons is then rather a particular consequence of this more general scenario.

In a first step we will provide arguments for the understanding of \tit{quantum space-time} (QST), that is, the microscopic structure which underlies our ordinary classical smooth space-time manifold ST, as a thermal system at each macroscopic point $x\in ST$. This will be done in the next section (some remarks in this direction have already been made in \cite{Requ1}. Early ideas can already be found in the work of Sakharov but without a thermal connotation (see e.g. \cite{Sak} and \cite{Visser}).
 We will exploit more recent findings in the foundations of modern quantum statistical mechanics (\cite{Popescu1},\cite{Lebo1},\cite{Lebo2}). Furthermore we will introduce the notion of the classical space-time metric $g_{ij}(x)$ as an \tit{order parameter field} and classical smooth space-time as an \tit{order parameter manifold}.

By these notions we mean the following. As briefly described in \cite{Requ1}, we assume that (Q)ST emerged as the consequence of a primordial phase transition PT. In this context a non-vanishing field of observables $g_{ij}(x)$ came into existence which was zero or rather vanishing before the phase transition happened in the primordial soup. That is, $g_{ij}(x)$ and ST represent parameters which indicate that a certain phase transition took place. Such a quantity is typically called an order parameter.

We start from the following assumption or observation: Classical space-time ST remains to be smooth on all ordinary scales of the quantum matter regime. So we conjecture that the typical scale where quantum gravity effects emerge is the \tit{ Planck scale}, that is
\beq   l_p=\hbar G/c^3\quad ,\quad t_p=\hbar G/c^5=l_p/c    \eeq
We hence assume that the typical quantum gravity degrees of freedom (DoF) live on the same scale. This implies that, as e.g. in hydrodynamics, there exist many quantum gravity DoF in the infinitesimal neighborhood of the macroscopic point $x\in ST$ in the classical space-time manifold ST. That is, we can assume that at each macroscopic point $x$ exists a microscopic subsystem in QST, consisting of many DoF. That these subsystems behave thermally will be proved in the next section. 

In a second step we will interpret the EEQ (sign convention as in \cite{Wheeler1}) 
\beq  G_{ab}:= R_{ab}-1/2\cdot R\cdot g_{ab}=\kappa \cdot T_{ab}\quad ,\quad \kappa=8\pi G/c^4   \eeq
in this thermodynamic context. That is, even if we assume that the global state of the microscopic QST is pure, the local state around some macroscopic point $x$ is a thermal ensemble (density matrix) due to its intricate entanglement with its environment. From the findings in the papers, mentioned above, we may then conclude that with overwhelming probability the local states are thermal.

As the local states around the points of the space-time manifold ST are thermal it seems reasonable to give also the EEQ a thermal interpretation. One should note that already Lorentz or Levy-Civita in 1916 or 1917 tried to interpret the EEQ as saying that the total energy of the universe is vanishing (cf. \cite{Pauli1})
\beq  (T_{ab}-\kappa^{-1}G_{ab})\equiv 0 \quad \text{hence}\quad \partial_a (T^a_b-
\kappa^{-1}G^a_b)\equiv 0 \quad \partial_a:=\partial/\partial x^a \label{EEQ}\eeq
Einstein provided an argument against such an interpretation (see \cite{Pauli1}) which seems to be sound at first glance, but one should take into account that at that time quantum theory was not yet fully developed (cf. e.g. our arguments, given in \cite{Requ1}).

In this paper we will only comment on the vast field of gravitational energy and its localization in general relativity (GR) in so far as it concerns the problems we want to analyze in the following. An interesting point of view is developed for example in \cite{Coop1}, where it is claimed that gravitational energy should be located at places where $T_{ab}\neq 0$. However, one generally assumes that non-vanishing curvature has something to do with gravitational energy. The latter one can of course be non-zero where $T_{ab}$ vanishes.

If the microscopic substructure , i.e. QST, of the classical smooth space-time ST is locally a thermal system, it should possess what one calls \tit{internal energy} in thermodynamics. As in \cite{Requ1} we assume the existence of an interaction between matter and space-time on the quantum level, that is, between QST and QM (quantum matter). The macroscopic ponderable matter and fields making up the right side of the EEQ consist of many microscopic DoF on the quantum scale. The same holds for the left side which describes a curvature property of ST.

All this leads us to the following conjecture:
\begin{conjecture} We assume that
\beq  T_{ab}=\kappa^{-1}G_{ab}  \eeq
is the heat influx at point x into the local thermal state of QST coming from the external QM system. 
\end{conjecture}
Note that this concerns only the gravitational effect of matter. Thus $\kappa^{-1}G_{ab}$ represents a form of \tit{heat tensor} which was discussed for material systems e.g. in \cite{Moeller1} or\cite{Eckart1}.
\begin{bem} We will later discuss in this paper the \tit{work terms} in this context,  \end{bem}

To complete our analysis of the thermal substructure of Gravitation or QST, we have firstly to deal with the \tit{non-tensorial} character of gravitational energy. Usually it holds 
\beq  \partial_a(\sqrt{-g}(T^a_b+t^a_b))\equiv 0  \eeq
with $t^a_b$ the \tit{pseudo tensor} of gravitational energy-momentum (brief discussions can be found in practically every book about GR, a particularly nice discussion is e.g. given in \cite{Xulu1}). In the canonical view this pseudo tensor property is frequently considered to be a deficit of the theory. In section (3) we reconsider the situation from a perhaps new vantage point.

It is a widespread belief that all concepts and notions which carry a physical meaning should be of tensorial character. It is however reasonable to analyze why tensorial behavior is considered to be of such a great relevance. If we have a manifold with tangent and cotangent spaces at each point, tensor fields can be built over tensor products of these spaces. Their importance is hence that they carry a geometric meaning, that is, tensors may be said to represent the same \tit{invariant} object independent of the choice of coordinate systems. In this context they have the important property that they cannot be transformed away (transformed to zero) by an appropriate choice of coordinate system. In particular, their transformation properties are relatively transparent.

On the other hand this philosphy cannot be exactly true as the wellknown \tit{connection coefficients} or \tit{Christoffel symbols} $\Gamma^I_{jk}$ are of great geometric significance and do not have a particularly simple transformation behavior. In our view, Weinberg rightly  remarks that there is nothing sacred about the tensor transformation law  (\cite{Wein1}). This holds the more so as physical quantities are not automatically elements of the multilinear algebra over tangent and/or cotangent space (which have essentially the meaning of velocity or momentum vectors).

In older text books physical tensorial quantities are hence introduced as tuples having simply a certain transformation behavior. We therefore should be prepared that there may exist physical quantities which do not transform as tensors. In the valuable and highly informative essay \cite{Rowe1} there is an interesting remark by Einstein, when asked by the Swiss student Humm, if gravitational energy should be tensorial or not. He thinks \tit{not}, referring for example to the different behavior of kinetic energy $T$ and potential energy $U$ under the Galilei group in
\beq  \partial_t (T+U)=0   \eeq

Another argument, in our view, would be a pendulum clock with friction which allows us to extract energy from the gravitational field in a non-inertial reference system, which is not possible in a freely falling inertial system. This latter example suggests the following conclusion:
\begin{ob} The equivalence principle shows that gravitational energy cannot be tensorial.  \end{ob}

We will discuss the deeper reasons for the possibility of non-tensorial behavior of various physical quantities in section (3) in a more systematic way, the main point being the effects of quantum theory on space-time and gravitational behavior. These quantum effects will lead to a subtle distinction between surface properties living over the smooth classical space-time manifold ST and the behavior in the underlying microscopic substructure denoted by QST. The former are expected to be tensorial, the latter typically not. The spontaneous symmetry breaking of \tit{diffeomorphism invariance} will play an important role in this context.

In the last section we discuss in more detail the relevance of diffeomorphism invariance and the corresponding conserved \tit{Noether current} for our investigation of the thermal character of the EEQ and GR in general. Wald studied the Noether current of diffeomorphism invariance in the framework of differential forms lying the main emphasis on integral representations (\cite{Wald2},\cite{Wald3},\cite{Wald4}). We will, on the other hand, rely more on the classical tensor analysis and variational approach, using mainly the non-integrated local point of view as it is e.g. also done in \cite{Pad1}.

But before we will do this, we will add some remarks about the quite detailed early work of Hilbert, Noether, Klein in this context (which, written in German, is perhaps not so widely known and which is carefully discussed in \cite{Rowe1}, for other aspects see also \cite{Sauer1}). These papers dealt mainly with a proper understanding of Hilbert's famous socalled \tit{energy vector}, which was derived from a quite intricate variational analysis of some invariant action functional and which seemed to puzzle not only the scientific community of that time quite a lot (\cite{Hilbert1},\cite{Klein1},\cite{Klein2},\cite{Noether1}. We find it remarkable that these long gone investigations anticipate in some sense more recent ones.

In Hilbert's analysis an arbitrary vector field $\xi^i$  was introduced in order to get a conserved current 
from diffeomorphism invariance. It was later understood (\cite{Noether1}) to be a necessary consequence of E.Noether's famous second theorem. We will analyze the remarkable behavior of this conserved current, give a physical interpretation of the vector field $\xi^i$, which is technically the generator of the  \tit{Lie derivative} and interpret the role of various terms occurring in the current as to their thermodynamic meaning and implications. This winds up to our observation that what Hilbert used to call the energy vector is rather an expression of the \tit{internal energy-momentum tensor} of the microscopic quantum gravitational substratum QST, underlying our smooth space-time manifold ST. 
\section{Quantum Space Time, QST, as a Thermal System}
As we have already remarked in the introduction, thermality usually enters the field of (quantum) gravity or general relativity (GR) via the existence of horizons, most notably in the paradigmatic case of black holes (BH). We want to argue in this and the following sections that, in our view, classical space time ST is the smooth, coarse grained hull, overlying QST, which is assumed to be a system having quantum micro structure, that is, consists of a complex network of a huge number of microscopic gravitational degrees of freedom (DoF).
\begin{assumption} We make the simplifying assumption that we may restrict ourselves to an intermediate energy scale where all DoF can be treated as quantum. Furthermore, we assume the gravitational DoF to live on a scale which is given by the Planck units.
\end{assumption}

From this we may infer the following. Compared to the Planck scale, the ordinary quantum scale on which our quantum matter is living, is many orders of magnitude larger. This is the reason why on the level of ordinary quantum theory space time can be treated as smooth and relatively invariable. On the other hand, we have briefly described in e.g. \cite{Requ1} how the microscopic gravitational DoF, living in QST cooperate so that as a consequence of a primordial phase transition PT the classical space time manifold ST with its points $x$ and the classical metric tensorfield $g_{ab}(x)$ do emerge, with $g_{ab}(x)$ playing a double role as allowing, on the one hand, to measure distances and time intervalls and, on the other hand, being the carrier of the gravitational field. 

As in \cite{Requ1} we introduce a Hilbert space $\mcal{H}^g$, in which the global quantum states, $\psi^g$, of QST are living. The microscopic elementary gravitational DoF are represented by their corresponding local Hilbert spaces $\mcal{H}^g_i,\, i\in \N$. The local bases in $\mcal{H}^g_i$ are denoted by $d_i^{\nu_i}$. Basis vectors in $\mcal{H}^g$ are then given by tensor product states
\beq  \psi^g_I=\bigotimes_i\, d_i^{\nu_i}\, ,\,I:=\{\nu_1,\nu_2,\cdots\}   \eeq
and a general pure quantum state in $\mcal{H}^g$ is
\beq  \psi^g=\sum_I\, c_I\psi_I^g     \eeq
\begin{bem} We note that these assumptions are not really necessary prerequisites to reach the following conclusions but will represent a convenient model to make the derivations more concrete and precise. \end{bem}

We have now to discuss what consequences can be inferred from the existence of the nonvanishing classical metric tensor field $g_{ab}(x)$ which lives over the classical space-time manifold ST and by the same token over the underlying microscopic substructure QST. Usually it is assumed that a quantum observable $\hat{g}_{ab}(x)$ exists on QST with
\beq g_{ab}(x)=<\hat{g}_{ab}(x)>:=(\psi^g\mid \hat{g}_{ab}(x)\psi^g)  \eeq
for some global state $\psi^g$ which corresponds as a quantum correlate to the classical ST.

This, at first glance, natural assumption leads however to a number of highly non-trivial consequences and has now to be discussed in more detail. We assumed for example in \cite{Requ1} that the existence respectively non-vanishing of the metric tensor field $g_{ab}(x)$ is the result of a primordial phase transition PT. In other words, the existence of a stable space-time distance concept is a non-trivial consequence of a particular physical process. Hence, following the tradition of e.g. quantum statistical mechanics as a paradigm, dealing with a great number of quantum DoF, we define:
\begin{defi} We call $g_{ab}(x)$ an order parameter field and the space-time manifold ST an orderparameter manifold, where its non-vanishing signals the existence of a transition to a greater structural order in the underlying microscopic substratum (see e.g. \cite{Requ1} or \cite{Requ2},\cite{Requ3} and earlier references given there).  \end{defi}

In our context, 
 where we have a double structure of an overlying classical smooth space-time manifold ST and an underlying microscopic quantum structure QST, both being correlated in a subtle way, we have to introduce some particular structural elements, that is, for example, the concept of \tit{macro observables}. As far as we know, this notion was for the first time introduced by v.Neumann in an ingenious but perhaps little kown paper (see \cite{Neum1}, the second part of \cite{Neum2} or \cite{Requ4}).

Due to the vastly different scales of, on the one hand, ordinary quantum matter QM and, on the other hand, quantum space-time QST, we can assume that to a \tit{macroscopic point} $x$ do belong a huge number of DoF in QST or \tit{local Hilbert spaces} $\mcal{H}_i$ which live in the infinitesimal neighborhood of $x$. We denote this situation by
\beq \mcal{H}_x:= \bigotimes_x\, \mcal{H}_i \sim [x]    \eeq
where $[x]$ denote the gravitational DoF in the infinitesimal neighborhood of $x$.
Furthermore, to a macroscopic value, $g_{ab}(x)$, we can choose a corresponding subspace $[g_{ab}(x)]$ in $\bigotimes_x\,\mcal{H}_i$ containing the local vectors $\psi^g(x)$ with
\beq  <\psi^g\mid \hat{g}_{ab}(x)\psi^g>=g_{ab}(x)  \eeq
\begin{defi} Following the tradition in quantum statistical mechanics, we call $[g_{ab}(x)]$ a phase cell and which is spanned by certain basis vectors $\psi_i^g(x)$ so that
\beq  \psi^g(x)=\sum\, c_i\psi^g_i(x)   \eeq
\end{defi}

Furthermore, in the class of global states $\psi^g$, we may select the subclass $[g_{ab}]$ so that 
\beq  <\psi^g\mid \hat{g}(x)\psi^g>=g_{ab}(x)\quad\text{for all}\; x\in ST  \eeq
\begin{bem} One should note that the various sets, $[x]$, or $[g_{ab}(x)]$, are not necessarily disjoint in QST. There may be a certain overlap. \end{bem}

We now come to a central point in our analysis. Given a global state $\psi^g$ in $\mcal{H}^g$, we can test it by sufficiently localized observables, that is, observables localized in an infinitesimal neighborhood of some macroscopic point $x$. That is, we may concentrate ourselves on state vectors from $\mcal{H}_x$ or $[g_{ab}(x)]$. More specifically, if we have a pure global state $\psi^g$ and test it with observables taken from $\mcal{B}(\mcal{H}_x)$, the bounded operators operating on $\mcal{H}_x$, we can represent the global state $\psi^g$ by a density matrix $\rho_{\psi}$, living over $\mcal{H}_x$ or $[g_{ab}(x)]$.
\begin{ob} It exists a density matrix $\rho_{\psi}$ over $\mcal{H}_x$ or $[g_{ab}(x)]$ with
\beq  <\psi^g\mid A(x)\psi^g>=Tr(A(x)\cdot\rho_{\psi})  \eeq
for observables  $A(x)\in\mcal{B}(\mcal{H}_x)$
\end{ob}

It is a relatively recent observation that much stronger results can in fact be derived which lead to the thermalization results, which we mentioned in the introduction. To our knowledge, early results in this direction were already derived by v.Neumann in \cite{Neum1}. More recently, these phenomena were analyzed in e.g. \cite{Popescu1},\cite{Lebo1}, or \cite{Lebo2}. In the context of \tit{decoherence by environment} see also \cite{Requ4}. Important tools in this connection are, on the one hand, the socalled \tit{concentration of measure phenomenon} and the Levy estimates (see for example \cite{Popescu1}, a more systematic discussion can be found in \cite{Ledoux}) and, on the other hand, the use of \tit{random (unit) vectors} (that is, quantum mechanical states lying on high-dimensional unit spheres) and socalled \tit{typical states}. We will not go into the complex technical details but restrict ourselves to an application of the results to our particular situation.

That is, we are given the classical order parameter field $g_{ab}(x)$, defining the class of microscopic quantum states $\psi^g\in [g_{ab}]$ with
\beq  <\psi^g\mid \hat{g}(x)\psi^g>=g_{ab}(x)\quad\text{for all} x\in ST  \eeq
We observed above that locally $\psi^g$ can be replaced by a trace operator or density matrix $\rho_{\psi}(x)$.

 The results derived in the above cited papers, in particular \cite{Popescu1}, now yields the following in our scenario. 
\begin{conclusion} For a global pure state $\psi^g\in[g_{ab}]$ it hold with very high probability that for local observables it is equivalent to the \tit{micro canonical ensemble}, i.e.
\beq  \Omega:=d^{-1}\cdot Tr(\mbf{1}\cdot)  \eeq
with d the dimension of $[g_{ab}]$ and $\mbf{1}$ the identity operator. The local state $\rho_{\psi}(x)$ at some arbitrary macroscopic point is equivalent to this $\Omega$ and is called the generalized canonical ensemble. In other words, this holds for a typical or randomly selected state. One should however note that $\Omega$ is a global state, $\rho_{\psi}(x)$ a local one.
\end{conclusion}
\begin{bem} In case the generalized canonical ensemble can be associated with some form of energy, the local states have the form of the true canonical ensemble known from thermodynamics. This is the reason why the corresponding local state can rightly be called a \tit{generalized canonical ensemble}. That is, the global micro canonical ensemble behaves locally as a canonical ensemble.  \end{bem}
\section{The Thermal Role of the Einstein Equation}
In section 2 we have argued that in an infinitesimal neighborhood of a macroscopic point $x\in ST$ quantum space-time (Q)ST has the characteristics of a local thermal system. We now want to introduce and discuss the various physical parameters which define and characterize such a thermal system. We begin with a new conceptual understanding of the EEQ. Originally the EEQ are viewed or conceived as a dynamic or evolution equation of the space-time continuum under the influence of (ponderable) matter and classical fields.

As mentioned in the introduction, much later Jacobson, Padmanabhan (and perhaps some other workers in the field) argued that the EEQ must hold due to the observed or conjectured thermal behavior near horizons, most notably, near the event horizon of a BH (see e.g. \cite{Jacob1},\cite{Jacob2},\cite{Pad1},\cite{Pad2},\cite{Pad3}). In the following we want to argue that the EEQ carry a direct thermal meaning in the bulk of ST, i.e., not necessarily near horizons and independent of the existence of horizons or singularities. This holds also for the concept of entropy, which is primarily understood in this context as a form of \tit{entanglement entropy}, typically arising near horizons. In our context it will have the ordinary thermodynamical meaning of number of configurational alternatives.

As we said in the introduction, already in the early days (1916,1917) e.g. Lorentz or Levi-Civita tried to give the EEQ a slightly different interpretation, which was however rejected by Einstein by an argument which was perhaps convincing at that time with quantum (field) theory still in its infancies (cf. \cite{Pauli1}). Lorentz and Levi-Civita speculated that by bringing the energy-momentum tensor, occurring on the rhs of the EEQ to the lhs, we get an expression which vanishes identically (see equation \ref{EEQ} in the introduction). They then conjectured that this expression can then be identified with the total (vanishing) energy of the universe, consisting of gravitational and matter energy.

Einstein argued that the vanishing of the total energy of the universe would not prevent the material systems and the gravitational energy to annihilate each other. However, meanwhile we know that exactly the opposite process might have happened in various models of the \tit{inflationary scenario} (creation from nothing, huge quantum fluctuations etc.). We discussed this zero energy universe idea in more detail in section 2 of \cite{Requ1}.

In \cite{Requ1} we introduced and described a similar but slightly different scenario. We argued that our universe, i.e. (quantum) matter, (Q)M, and (quantum) space-time, (Q)ST, emerged both as a result of a primordial phase transition PT from a more primordial phase, QX:
\beq  QX\underset{PT}\rightarrow (Q)ST+(Q)M  \eeq
While the internal energy of QST is lowered compared to the original phase QX as the result of an \tit{order transition} by an amount $\Delta E$, this same amount is transferred to QM. That is,
\beq E(QX)=E(QST)+E(QM)  \eeq
 
In the language of Hilbert spaces we describe the structure of our universe as a pure state in the tensor product
\beq  \psi\in\mcal{H}=\mcal{H}^g\otimes\mcal{H}^M   \eeq
with 
\beq \psi=\psi^g\otimes \psi^M  \eeq
being a state where the gravitational and the matter state do exist independently from each other. This may be an approximation of a situation where e.g. BH's are absent. The BH situation was discussed in \cite{Requ1}.

A thermal system should have an \tit{internal energy}. In the introduction we formulated the following conjecture:
 \begin{conjecture} We assume that
\beq  T_{ab}(x)=\kappa^{-1}G_{ab}(x)  \eeq
is the heat influx at point x into the local thermal state of QST coming from the external QM system. 
Note that this concerns only the gravitational effect of matter. Thus $\kappa^{-1}G_{ab}$ represents a form of \tit{heat tensor} which was discussed for material systems e.g. in \cite{Moeller1} or\cite{Eckart1}.     
\end{conjecture}
That is, the EEQ is essentially a statement about the total heat influx at point $x\in ST$ into the infinitesimal neighborhood of $x$ in QST. We now want to provide arguments why we think this is indeed the case.

In \cite{Requ5} we studied thermodynamics in the regime of special relativity. While in this field there do exist a variety of different approaches, in the approach we favored, i.e. with temperature being the zero component of a contravariant four-vector, 
\beq  T=\gamma\cdot T_0\quad ,\quad \gamma=(1-u^2/c^2)^{-1/2}  \eeq
with $\gamma$ the Lorentz factor, $T_0$ the rest temperature in a comoving coordinate system, $u$ the 3-velocity, $T$ the temperature in the laboratory frame, we get the following interesting transformation property for the heat influx into the system:
\beq \delta Q=\gamma\cdot \delta Q_0   \eeq
that is, in contrast to internal energy and work, the heat influx transforms as the zero component of a contravariant 4-vector. That is, it transforms covariantly (cf. in particular section 4.4 of \cite{Requ5}).

Another paradigm is BH physics and the first law of BH-thermodynamics. In geometric units (c=G=1) it reads:
\beq  \delta M=(8\pi)^{-1}\cdot \kappa\delta A+\Omega\delta J   \eeq
with $\kappa$ the \tit{surface gravity}, $\delta A$ the change of area, $\Omega$ angular velocity of horizon, $\delta J$ change  of angular momentum, $\delta M$ the change of mass in the center. The identification reads:
\beq  \delta S_{BH}=\delta A/4\, , \, T=\kappa/2\pi\, ,\, \delta M\,\text{change of internal energy} \eeq

It is perhaps surprising that heat and entropy occur at such a prominent place in this field while the, at first glance, more natural work contributions seem to be missing. We now try to explain this on a perhaps more fundamental level. We already briefly mentioned in the introduction that, in our view, space-time consists of at least two levels, a smooth surface structure, ST, with its tangential and cotangential structure and an underlying microscopic quantum structure, QST. This compound system is assumed to have emerged via a primordial phase transition, PT, together with a (quantum) matter component, QM.

This phase transition was a transition to a greater order, that is, (Q)ST carries an extra space-time structure compared to the more primitive phase, QX, given, for example, by the existence (or non-vanishing) of the metric tensor field, $g_{ab}(x)$, having the character of an order parameter field, and in the underlying QST the existence of certain phase cells, $[g_{ab}(x)]$, containing the quantum states of the universe, leading to the same macroscopic metric tensor field in ST, i.e. $g_{ab}(x)$. Therefore we can conclude:
\begin{conclusion} The primordial phase transition, PT, is accompanied by a spontaneous symmetry breaking, SSB, of diffeomorphism invariance (or covariance). That is, the more symmetric but less ordered phase, QX, goes over into the phase (Q)ST, which has lesser symmetry but higher order.
\end{conclusion}
\begin{bem} These points were already discussed in \cite{Requ2} and \cite{Requ3}.  \end{bem}.

What are the consequences of these observations for our question of non-covariance of e.g. gravitational energy and related questions? As we said above, we have a two storey structure of space-time. We argue that the smooth surface structure ST (i.e. the macroscopic part) supports observables which are diffeomorphism invariant. The SSB, on the other hand, is located in the micro structure of the quantum regime QST. That is, we assume that the existence of quantum excitations, vacuum fluctuations and the emergence of the space-time structure $[g_{ab}(x)]$ is responsible for the breaking of diffeomorphism invariance. We conjecture the following:
\begin{conjecture} We assume that quantities which live over the smooth macroscopic manifold ST are diffeomorphism covariant, i.e. behave tensorial. On the other hand, quantities which rather express properties of the microscopic underlying quantum structure QST we assume to behave non-tensorial in general.
\end{conjecture}
\begin{koro} As we already explained in the introduction, since the scale on which QST is living is many orders of magnitude finer than even the regime of ordinary quantum matter, the latter should also behave covariantly in general.
\end{koro}
\begin{conclusion} As the energy-momentum tensor, $T_{ab}(x)$, lives essentially over the classical manifold ST, the heat influx, $\kappa^{-1}\cdot G_{ab}(x)$, is by the same token also tensorial or covariant. On the other hand, the part of the relativistic energy-momentum (pseudo) tensor, $t_{ab}(x)$, describing the gravitational energy contained in QST, is non-tensorial because it describes the contributions to gravitational energy which are contained in the underlying network of gravitational DoF. making up the microscopic structure of QST.
\end{conclusion}
\section{The Concept of Gravitational Temperature}
On general grounds we know that our local systems must have a temperature. From what we have said above, the local gravitational system around the point $x\in ST$ has an internal energy $U(x)$ and an entropy $S(x)$. In section 2 we showed that the system at $x\in ST$ can be considered as a \tit{generalized canonical ensemble}. This entails that we can define its v.Neumann entropy 
\beq  S(x):= -Tr\,(\rho_x\cdot \ln\, \rho_x)=-\sum_i\, p_i\ln\,p_i    \eeq
with $p_i$ the probabilities of the individual states $\psi_i(x)$. Therefore we can (in principle) calculate 
\beq T(x)=\partial U(x)/\partial S(x)\;\text{with}\; V(x)\;\text{held fixed} \eeq 
($V(x)$ some infinitesimal volume element around $x$).

We will complete this somewhat abstract definition of gravitational temperature by two different more concrete approaches in this context. Firstly, we discuss the Tolman-Ehrenfest approach (cf. e.g. \cite{Tolman1},\cite{Tolman2}) which was originally introduced for thermal material systems which live in ST. For convenience of the reader we will briefly recapitulate our own approach which we gave in \cite{Requ6}.
\begin{bem} In order that we can apply the Tolman-Ehrenfest results to purely gravitational systems, 
that is, systems not consisting of ordinary material constituents, we must at first motivate that these systems have a thermal character and consist of microscopic constituents. This is what we have done in this paper up to now.
\end{bem}
 
The Tolman-Ehrenfest result is the following:
\begin{ob} In thermal equilibrium in a static gravitational field we have for an isolated system
\beq T(x)\cdot\sqrt{-g_{00}(x)}= const   \eeq
I.e., in contrast to the non-relativistic regime (cf. sect.2), there exists in general a temperature gradient in a system being in thermal equilibrium in the relativistic regime.
\end{ob}
To derive this result we use the \tit{entropy maximum principle}. We shall use, for reasons of simplicity, the weak field expansion of the gravitational field. With $\phi(x)$ the Newtonian gravitational potential we have
\beq \sqrt{-g_{00}}=(1+2\phi/c^2)^{1/2}\approx 1+\phi/c^2   \eeq
\begin{bem} Note that the gravitational potential is negative and is usually assumed to vanish at infinity.
\end{bem}
We assume an isolated macroscopic system to be in thermal equilibrium in such a static weak gravitational field. Its total entropy and internal energy depend on the gravitational field $\phi(x)$. We now decompose the large system into sufficiently small subsystems so that the respective thermodynamic variables can be assumed to be essentially constant in the small subsystems. As the entropy is an \tit{extensive} quantity, we can write
\beq S(\phi)=\sum_i S_i(E^0_i,V_i,N_i)   \eeq
where $E^0_i$ is the thermodynamical \tit{internal energy}, not including the respective \tit{potential energy}.
\begin{ob} It is important that in the subsystems the explicit dependence on the gravitational potential has vanished. The entropy in the subsystems depends only on the respective thermodynamical variables, the values of which are of course functions of the position of the respective subsystem in the field $\phi(x)$.
\end{ob}

At its maximum the total entropy is constant under infinitesimal redistribution of the internal energies $E^0_i$ with the total energy and the remaining thermodynamic variables kept constant. We now envisage two neighboring subsystems, denoted by (1) and (2). To be definite, we take $\phi(2)\geq \phi(1)$. We now transfer an infinitesimal amount of internal energy $dE_2^0$ from (2) to (1) (note, it consists of pure heat as for example the particle numbers remain unchanged by assumption!). As heat has weight relativistically it gains on its way an extra amount of potential energy.
\begin{ob} By transferring $dE_2^0$ from (2) to (1) we gain an additional amount of gravitational energy 
\beq dE_2^0\cdot\Delta\phi/c^2 \quad ,\quad \Delta\phi=\phi_2-\phi_1   \eeq 
It is important to realize that this gravitational energy has to be transformed from mechanical energy into heat energy or, rather, internal energy and reinjected in this form into system (1) (for example by a stirring mechanism acting on system (1) and being propelled by the quasistatic fall of the energy  $dE_2^0$).
\end{ob}
The energy balance equation now reads
\beq dE_1^0=dE_2^0+dE_2^0\cdot\Delta\phi/c^2=dE_2^0(1+\Delta\phi/c^2)  \eeq 
We then have (with $dS_1=-dS_2$ in equilibrium)
\beq T_1^{-1}=dS_1/dE_1^0=-dS_2/-(dE_2^0(1+\Delta\phi/c^2))=T_2^{-1}\cdot (1+\Delta\phi/c^2)^{-1}   \eeq
that is
\begin{conclusion} It holds
\beq T_1=T_2(1+\Delta\phi/c^2)\quad ,\quad \Delta\phi:=\phi_2-\phi_1\geq 0   \eeq
\end{conclusion}
\begin{ob} The subsystem (1), having a potential energy being lower than (2), has a higher temperature.
\end{ob}

We can give the above relation another more covariant form. In the approximation we are using it holds:
\beq \frac{\sqrt{1+2\phi_2/c^2}}{\sqrt{1+2\phi_1/c^2}}=\frac{1+\phi_2/c^2}{1+\phi_1/c^2}=\frac{1+(\phi_1+\Delta\phi)/c^2}{1+\phi_1/c^2}=1+\Delta\phi/c^2   \eeq
\begin{conclusion}[Covariant form] It holds
\beq T_1\cdot\sqrt{-g_{00}(1)}=T_2\cdot\sqrt{-g_{00}(2)}=const    \eeq
\end{conclusion}
The non-infinitesimal result follows by using a sequence of infinitesimal steps.

It is perhaps useful to derive the above result in yet another, slightly different, way. In case $\phi(x)$ vanishes at infinity we can apply the entropy-maximum principle as follows. We extract the energy (bringing it to infinity)
\beq dE_2=dE_2^0+dE_2^0\cdot \phi_2/c^2   \eeq
from subsystem (2) and reinject the energy
\beq  dE_2=dE_1=dE_1^0+dE_1^0\cdot \phi_1/c^2  \eeq
from infinity into (1). We get
\beq dE_1^0=dE_2^0\cdot\frac{1+\phi_2/c^2}{1+\phi_1/c^2}  \eeq
and
\beq   T_1^{-1}=dS_1/dE_1^0=T_2^{-1}\cdot\left(\frac{1+\phi_2/c^2}{1+\phi_1/c^2}\right)^{-1}   \eeq
hence
\beq  T_1\cdot (1+\phi_1/c^2)=T_2\cdot (1+\phi_2/c^2)   \eeq

As an example one may mention the Rindler/Unruh space-time. We have in Rindler coordinates
\beq g_{00}=-\xi^2\; ,\; a=\xi^{-1}\; ,\; T=a/2\pi   \eeq
i.e.
\beq T\sqrt{-g_{00}}=2\pi^{-1}=const   \eeq
\begin{bem} One should mention that this method to go to $\infty$ does make only sense if the thermal system extends to $\infty$ as well.   \end{bem}

In a second approach we will employ the Fulling-Unruh observation of thermalisation in accelerated frames of reference (local Rindler frames were introduced by for example Padmanabhan in e.g. \cite{Pad2}). We will however use it in a slightly different way compared to Jacobson or Padmanabhan. For reasons of convenience we wil choose a static space-time ST and restrict ourselves to a sufficiently small neighborhood of some arbitrary point $x\in ST$.

We choose a local inertial frame (LIF) at $x$ with Lorentz-orthonormal coordinates $(X,T)$ so that $x$ has the coordinates $(0,0)$. We assume that the LIF moves along a certain geodesic through $x$. We place a thermometer system at the point $x$ which is gauged so that it yields the result zero in a LIF. However, we assume that the thermometer is in principle sensitive against the thermal excitations of the underlying microscopical gravitational system. The thermometer now experiences the gravitational field $g_{ab}(x)$ at the macroscopic point $x\in ST$.

Now, relative to the LIF (moving on a geodesic through $x$) and employing the inertial coordinates $(X,T)$ an observer at the point $x$ (and the thermometer) experiences an acceleration $a(X,T)$ (the detailed numerical calculations, which are a little bit intricate, can be found in \cite{Moeller1} exercise in section 9.6). One should however note that everything we have said holds only locally. But locally, near the origin, we can switch from inertial coordinates $(X,T)$ to socalled Rindler coordinates $(x_R,t_R)$.
\beq T=x_R\cdot \sinh\,(\kappa t_R)\; ,\; X=x_R\cdot\cosh(\kappa t_R)   \eeq
with $\kappa$ the corresponding proper acceleration.

In full Rindler space, i.e. the right wedge $W_R$, we know from the work of Fulling (\cite{Full1}) and Unruh (\cite{Un1}) that the thermometer will detect thermal Rindler modes. In our case the situation is only locally, i.e. in a neighborhood of $x$ or $(X,T)=(0,0)$, Rindler like. Therefore we cannot expect to have such fully developed Rindler modes. However we know already that ST at $x$ or $(X,T)=(0,0)$ behaves microscopically as a thermal system, that is, as a socalled generalized canonical ensemble, hence carrying a distribution of local thermal excitations. Therefore we make the conjecture:
\begin{conjecture} The observer at $x$ or $(X,T)=(0,0)$ will detect a certain thermal excitation spectrum, consisting of certain quasi particles (which may be approximations of corresponding Rindler modes), leading to a corresponding local temperature.
\end{conjecture}
\section{Diffeomorphism Invariance and its Conserved Noether Current}
In section 3 we discussed the consequences of the possible SSB of diffeomorphism invariance due to the emergence of quantum DoF in the underlying micro structure of ST, i.e. QST as a result of the primordial phase transition PT. We argued that we may have two classes of observables, the one class of covariant observables which live above the macroscopic space-time manifold ST and another class, which we assume to consist of quantities which rather describe properties of the underlying micro structure within QST. We argued that the elements of this latter class need not behave covariantly under geometric transformations of the surface structure ST. A prominent example in this context is the concept of gravitational energy. Another important role will be played by the \tit{work terms} which occur in the complete conserved Noether current we will derive below. Their covariance comes about by exploiting the vector field $\xi^i(x)$ which enters via the Lie derivative, which is a covariant operation.

But befor we enter into the technical details, we want to recapitulate what we said in the introduction concerning the work of Hilbert, Noether, Klein etc. (\cite{Hilbert1},\cite{Klein1},\cite{Klein2},\cite{Noether1}). We were surprised to see that Hilbert essentially got after a long and intricate calculation a conservation law for what he considered to be the gravitational \tit{energy vector}. This energy vector contained an arbitrary vector field $\xi^i$, the deeper role of which was not really understood and appreciated at that time, while technically it is a consequence of E.Noether's \tit{second theorem}. In our view Hilbert already performed calculations which were much later repeated in a similar context, being apparently unaware of the earlier results. Below we will explain the physical role of the arbitrary vector field $\xi^i$ while its mathematical role is clear, it is simply the vector field which generates the diffeomorphism group and occurs in the Lie derivative. In traditional physicist's notation:
\beq  x^i\to \overline{x}^i=x^i+\epsilon\xi^i(x)  \eeq
 with $\epsilon$ infinitesimal. These intensive investigations and discussions performed in the Hilbert group were carefully studied in two beautiful essays by Rowe and Sauer (\cite{Rowe1},\cite{Sauer1}).

We give now a brief description of the sequence of necessary steps which lead to the form of conserved Noether current deriving from the assumption of diffeomorphism invariance. We will perform most of these steps in an appendix. Our main motivation for this is a consequence of the observation that in many representations of this stuff for example important and \tit{nonvanishing} boundary terms are frequently dropped which then results in only partial results.
\begin{bem}If for example certain boundary terms are dropped one gets only the \tit{contracted Bianchi identity} instead of the full conserved current. This is in our view dangerous because in contrast to ordinary variations diffeomorphisms typically need not vanish at infinity. Furthermore, we will show that variations, having a local support, lead to results which differ in some important respects.
\end{bem}

In the following we restrict ourselves to a variation of the Hilbert-Einstein action
\beq S[g_{ab}]=\int\,R(g_{ab})\cdot \sqrt{-g}\,d^4x   \eeq
with $R$ the scalar curvature
\beq R:=R_a^a\;\text{with}\;R_{ab}\;\text{the Ricci tensor}\; R_{ac}:=R_{abc}^b  \eeq
the rhs being the Riemann curvature tensor. $g$ is the determinant of the metric tensor. $R$ is a scalar and $\sqrt{-g}\,d^4x$, the \tit{canonical volume element}, is an invariant under coordinate transformations or diffeomorphisms. Hence $S$ is invariant under diffeomorphisms
\beq \phi_{\lambda}:\,x\to\,x(\lambda)\;\text{or infinitesimal:}\;x\to\bar{x}=x+d\lambda\,\xi(x) \eeq
where the vector field $\xi(x)$ induces the flow of the diffeomorphism group $\phi_{\lambda}$.

If $T(x)$ is some arbitrary tensor field, we construct an $\lambda$-dependent tensor field in the following way. We shift back the tensor at $x(\lambda)$ to the point $x$
\beq T_{\lambda}(x):=\phi^*_{-\lambda}(T(x_{\lambda}))   \eeq
 with $\phi^*_{-\lambda}=(\phi^*_{\lambda})^{-1}$ the map, induced by $\phi_{\lambda}:x\to x(\lambda)$. Now all these $T_{\lambda}(x)$ are defined at the same point $x$ and we can take the derivative with respect to $\lambda$ at $\lambda=0$. We get
\beq 0=d\,S[g_{ab}]/d\,\lambda=\int\,d/d\lambda\,\mcal{L}(g_{ab}(x;\lambda)\,d^4\,x  \eeq
where in the following the derivative is taken always at $\lambda=0$ and $\mcal{L}$ is the scalar density $R(g_{ab})\cdot \sqrt{-g}$.
\begin{bem}Technical details can be found, for example, in \cite{Wald5}, Appendix C. \end{bem}

\begin{ob}The derivative with respect to $\lambda$ at $\lambda=0$ is nothing but the Lie derivative, $\mcal{L}_{\xi}$, of the tensor $T$ or more general geometric objects. As with ordinary derivatives, the Lie derivative obeys the Leibniz rule. \end{ob}
\begin{defi}In the following we abreviate 
\beq \mcal{L}_{\xi}(T)(x):=\delta T(x)  \eeq
that is, in particular
\beq d/d\lambda\,g_{ab}(x;\lambda)=\mcal{L}_{\xi}\, g_{ab}(x)=\delta g_{ab}(x)  \eeq
\end{defi}

It is an important observation that, as the Lie derivative is defined for general differentiable manifolds, it is independent of the concept of covariant derivative. It can hence be expressed with the help of, on the one hand, partial derivatives, on the other hand, expressed by means of an arbitrary covariant derivative operator. For example with the help of the covariant derivative, induced by $g_{ab}(x)$:
\beq \mcal{L}_{\xi}\,g_{ab}=\nabla_a\xi_b+\nabla_b\xi_a=\xi^c\partial_c g_{ab}+g_{cb}\partial_a\xi^c+g_{ac}\xi^c  \eeq

While the Lie derivative for tensor fields is canonically given via $\phi^*_{-\lambda}(T(x_{\lambda}))$ as
\beq \mcal{L}_{\xi}(T)(x):=\lim_{\lambda\to 0}\;(\phi^*_{-\lambda}(T(x_{\lambda}))-T(x))/\lambda \eeq
one has to say some words in the case of e.g. densities.
\begin{bem}A scalar density like $\mcal{L}(x)$ becomes an invariant if multiplied by the volume element $d^4x$. If we have a scalar density at point $x(\lambda)$, its translate back to $x$ has to be again a scalar density so that e.g. $\mcal{L}\,d^4(x_{\lambda})$ remains invariant. 
\end{bem}
What we have said in the remark, allows us to fix the necessary transformation properties and define the Lie derivative of a scalar density.
\begin{lemma}We have
\beq \mcal{L}_{\xi}(\sqrt{-g}\cdot R)=-\sqrt{-g}\nabla_a(R\cdot\xi^a)=-\partial_a(R\cdot \xi^a\cdot\sqrt{-g})  \label{55} \eeq
\end{lemma}

We are now going to describe the strategy which will lead to the derivation of a conserved current, following from diffeomorphism invariance. We shall however not follow the perhaps more obvious strategy to exploit the above integral expression for $S[g_{ab}]$ but will use a local approach as it is for example done in \cite{Bjorken} in the case of the ordinary energy-momentum tensor conservation law following from 4-translation invariance in quantum field theory and also used in \cite{Pad1}. Our approach will consist of essentially two steps.

i) We take the Lagrange density $\mcal([g_{ab}(x)]$ which is a lengthy expression in $g_{ab}(x)$ and its first and second partial derivatives and, using the Leibniz rule and that partial derivatives and Lie derivative do commute, represent its Lie derivative as a long expression consisting ultimately terms containing the Lie derivative of the basic building blocks $g_{ab}(x)$ which we derived above. Note that we use the abbreviation
\beq d/d\lambda\, g_{ab}(x;\lambda)=\mcal{L}_{\xi}(g_{ab}(x)=:\delta g_{ab}(x)  \eeq
Diffeomorphism invariance enters in the way that $\mcal{L}[g_{ab}]$ does not explicitly depend on the coordinates but only via the field $g_{ab}$ and its derivatives. That is, this approach exploits the structural form of the Lagrange density.

ii) In a second step we simply directly calculate the Lie derivative of the density $R\sqrt{-g}$ as described in formula (\ref{55}). We then equate the numerically identical expressions and bring them on the same side, thus getting a vanishing expression which can be written as a conserved current. 

Now, using formula \ref{62} in the appendix and the contracted Bianch identity, we get for the variation of $\mcal{L}[g_{ab}]$:
\beq \delta\mcal{L}[g_{ab}]=\sqrt{-g}\cdot\nabla^a(v_a+2G_{ab}\xi^b)  \eeq
with
\beq v^a=\nabla^b(\nabla^a\xi_b+\nabla_b\xi^a)-g^{cd}\nabla^a(\nabla_c\xi_d+\nabla_d\xi_c)  \eeq
(some technical remarks:
\beq \nabla^a:=g^{ab}\nabla_b\quad\text{and}\quad \nabla^bg^{ac}=0=\nabla_bg_{ac}  \eeq
for the Levi-Civita connection).

Now using in the second step the direct Lie derivative of the Lagrange density $R\cdot\sqrt{-g}$, which we already wrote down above (see formula \ref{55}), we arrive at the conserved Noether current.
\begin{satz}The Lie derivative of $R\cdot\sqrt{-g}$, derived in the two ways, described above, yields the conserved Noether current
\beq 0=\nabla_a(R\xi^a+2G^{ab}\xi_b+v^a)  \eeq
with
\beq v^a=\nabla^b(\nabla^a\xi_b+\nabla_b\xi^a)-g^{cd}\nabla^a(\nabla_c\xi_d+\nabla_d\xi_c)  \eeq
\end{satz}
We can rewrite this formula to get a slightly different result. With
\beq 2G^{ab}\xi_b=2R^{ab}\xi_b-Rg^{ab}\xi_b\quad,\quad g^{ab}\xi_b=\xi^a  \eeq
we get
\begin{koro}A variant of the above result is
\beq 0=\nabla_a j^a\quad ,\quad j^a=(2R^{ab}\xi_b+v^a)  \eeq
\end{koro}
\section{The Thermal Meaning of the Conserved Noether Current}
We want now to come back to the interpretation of the conserved Noether current we have derived above. We mentioned in the introduction that Hilbert and his colleagues had great difficulties to understand the role the arbitrary vector field $\xi^i(x)$ is playing in this expression. As we have an energy-momentum tensor $T_{ab}$, playing a fundamental role in the theory, the occurrence of a conserved energy vector (as Hilbert liked to call it) like our $j^i(x)$ was puzzling. 

In this context we remind the reader what we said in section 3 concerning the role of tensorial or covariant quantities compared to non-tensorial or non-covariant quantities, the former referring to properties of the smooth classical surface structure ST, the latter ones referring to the underlying quantum mechanical micro structure QST. If we assume the vector field $\xi^i(x)$ is chosen to be time like, we can associate it with the orbits of observers or measuring devices. Then expressions like 
\beq T_{ab}(x)\xi^b(x)\quad \text{or}\quad G_{ab}(x)\xi^b(x)   \eeq
represent (energy) flows as observed or measured by the respective moving observers. That is, they have an objective (geometric) quality. In this sense they should have a covariant tensorial character and therefore can create a covariant conserved current.
\begin{ob}The (time like) vector fields $\xi^i(x)$ can be assumed to be the orbits of observers, floating through space-time.  \end{ob}

In section 3 we argued that
\beq G_{ab}(x)=\kappa\cdot T_{ab}(x)  \eeq
is a statement about gravitational heat energy influx at macroscopic space-time point $x$. Correspondingly the first part of the conserved Noether current describes the flow of heat energy contributing to the internal energy of the gravitational system. We now analyze the physical meaning of the second part $v^i(x)$.
\beq  v^a=\nabla^b(\nabla^a\xi_b+\nabla_b\xi^a)-g^{cd}\nabla^a(\nabla_c\xi_d+\nabla_d\xi_c)  \eeq
If $\xi^i(x)$ is a Killing vector field, that is, if it induces a symmetry of the metric tensor, 
\beq \mcal{L}_{\xi}g_{ab}(x)=0\quad\text{or}\quad \nabla_a\xi_b+\nabla_b\xi_a=0  \eeq
and using 
\beq \nabla_b\xi^a=\nabla_b(g^{ac}\xi_c)\; ,\; \nabla^a\xi_b=g^{ac}\nabla_c\xi_b\; ,\; \nabla_bg^{ac}=0  \eeq
we get:
\begin{satz}If $\xi^i(x)$ is a Killing vector field we have $v^i(x)=0$.  \end{satz}

Since in the preceding sections we argued that QST is a thermal system, possessing at each macroscopic space-time point $x$ the local state functions \tit{internal energy} and \tit{entropy} as well as a notion of local heat influx given by $T_{ab}(x)$ or $G_{ab}(x)$, it suggests itself to regard the vector field $v^i(x)$, which contains mainly contributions built from the metric, $g_{ab}(x)$, and the vector field $\xi^i(x)$ as work terms.
\begin{conjecture}We assume that the vector field $v^i(x)$ contains the work terms of our gravitational system QST. It contains essentially the effects of 4-volume changes (compressions and decompressions).  \end{conjecture}
\begin{ob}This interpretation is supported by the above result that $v^i(x)=0$ for Killing vector fields inducing geometric symmetries, $\mcal{L}_{\xi}g_{ab}=0$, i.e., essentially no 4-volume changes.  \end{ob}

In the introduction we mentioned the work of Sakharov. \tit{Induced Gravity} is assumed to result from the deformation of the structure of vacuum fluctuations by curvature. Our above working philosophy is a related one. Locally compressing or decompressing QST affects the local level structure of the thermal system at point $x$ and thus is a kind of work, done at the system. 

We would like to make some remarks concerning a point which irritates or irritated many researchers. It is frequently argued that mere coordinate transformations can completely alter the numerical values  of quantities like energy or work, even make them vanish in case we are dealing with non-tensor quantities. We dealt already with such problems in \cite{Requ4} in the context of special relativity (SR), in particular if volume changes due to Lorentz contraction have to be included in thermodynamic work terms.

Some researchers have the attitude to consider Lorentz contraction as not being real, whatever that actually means. We think Pauli in \cite{Pauli1} sect.5 made this point particularly clear. He argues that the atomic physics, underlying the contraction of a measuring rod is complicated but has to obey Lorentz covariance. Therefore Lorentz contraction, in his view, is, on the one hand, an objective process, but, as it is at the same time a result of Lorentz symmetry, it can as well be explained with the help of the general Lorentz invariance of SR.

The same is true, in our view, in the case of curvature effects on the micro structure of QST. Furthermore, we can choose at each space-time point $x$ a local geodesic coordinate system in which SR does hold, thus establishing the close relatedness of SR and GR. What concerns pure coordinate transformations, there exist basically two possibilities. On the one hand, they may be related to concrete changes of reference systems to which our above remarks do apply. On the other hand, they may not be implementable by concrete reference systems. In that case we may assume that transformation behavior should be regarded as a consequence following from consistency reasons.

\section{Appendix: The Conserved Noether Current}
We begin with the calculation of the variation of $R[g_{ab}]\cdot\sqrt{-g}$, that is, reducing $\delta(R[g_{ab}]\cdot\sqrt{-g})$ to an expression which contains only terms like $\delta g_{ab}$, remembering that $\delta$ denotes the Lie derivative or $d/d\lambda$ at $\lambda=0$. Furthermore we use the above expression for the Lie derivative
\beq \mcal{L}_{\xi}\,g_{ab}=\nabla_a\xi_b+\nabla_b\xi_a=\xi^c\partial_c g_{ab}+g_{cb}\partial_a\xi^c+g_{ac}\xi^c  \eeq
As $R$ is a relative complex expression in the variables $g_{ab}$ and its first and second partial derivatives, the calculations are, as is often the case in this context, lengthy and a little bit intricate.

We have (see \cite{Wald5} p.453)
\beq \delta(R\cdot \sqrt{-g})=\sqrt{-g}(\delta R_{ab})g^{ab}+\sqrt{-g}R_{ab}\delta g^{ab}+R\delta(\sqrt{-g})  \eeq
with $R_{ab}$ the Ricci tensor. Furthermore (\cite{Wald5} p.185)
\beq g^{ab}\delta R_{ab}=\nabla^av_a\quad v_a=\nabla^b(\delta g_{ab})-g^{cd}\nabla_a(\delta g_{cd})  \eeq
and (see \cite{Wald5} p.453, \cite{Pauli1} section 23)
\beq \delta(\sqrt{-g})=-1/2\cdot\sqrt{-g}g^{ab}\delta g_{ab}=+1/2\cdot \sqrt{-g}g_{ab}\delta g^{ab}  \eeq
\beq (0=\delta(g^{ab}g_{ab})=g^{ab}\delta g_{ab}+\delta g^{ab}g_{ab})  \eeq
\begin{conclusion}The above formulas yield
\beq \label{62} \delta\mcal{L}(g_{ab})=(R_{ab}-1/2Rg_{ab})\delta g^{ab}\sqrt{-g}+\nabla^av_a\sqrt{-g}  \eeq
\end{conclusion}
\begin{ob}$G_{ab}=(R_{ab}-1/2Rg_{ab})$ is called Einstein tensor with $G_{ab}=8\pi\cdot T_{ab}$. It fulfills the contracted Bianchi identity $\nabla^aG_{ab}=0$ which can also be derived from diffeomorphism invariance if one neglects boundary terms (see \cite{Feynman1} p.138). 
 \end{ob}

\end{document}